\def\propcom{$ \mathbf k =$ ($\frac1 2  \, \,0  \, \,\frac1 2$) }
\def\propinc{$ \mathbf k =$ ($\delta \, \,0 \, \, \frac1 2$) }
\documentclass[twocolumn, prb,superscriptaddress]{revtex4}       %  RevTex style article
\usepackage{color,graphicx}                               % Includes figure files
\usepackage{amssymb}                                      % Provides various mathematical symbols
\usepackage{lineno}                                            % Adds line numbers

\begin{document}
%--------------------------------------------------------------------------------------------------------------------------------------------------------
%  Front matter
\title{Magnetic-crystallographic phase diagram of superconducting parent compound Fe$_{1+x}$Te}

\author{E. E. Rodriguez}
\affiliation{NIST Center for Neutron Research, NIST, 100 Bureau Dr., Gaithersburg, MD 20878}
\author{C. Stock}
\affiliation{NIST Center for Neutron Research, NIST, 100 Bureau Dr., Gaithersburg, MD 20878}
\affiliation{Indiana University, 2401 Milo B. Sampson Lane, Bloomington, IN, 47408}
\author{P. Zajdel}
\affiliation{Division of Physics of Crystals, Institute of Physics, University of Silesia, Katowice, 40-007, Poland}
\author{K. L. Krycka}
\affiliation{NIST Center for Neutron Research, NIST, 100 Bureau Dr., Gaithersburg, MD 20878}
\author{C. F. Majkrzak}
\affiliation{NIST Center for Neutron Research, NIST, 100 Bureau Dr., Gaithersburg, MD 20878}
\author{P. Zavalij}
\affiliation{Department of Chemistry, University of Maryland, College Park, MD 20742}
\author{M. A. Green}
\affiliation{NIST Center for Neutron Research, NIST, 100 Bureau Dr., Gaithersburg, MD 20878}
\affiliation{Department of Materials Science and Engineering, University of Maryland, College Park,
MD 20742}

%
%--------------------------------------------------------------------------------------------------------------------------------------------------------
%  Abstract

\begin{abstract} Through neutron diffraction experiments, including spin-polarized measurements, we find a collinear incommensurate spin-density wave with propagation vector $ \mathbf k = $ ($0.4481(4) \, \,0 \, \, \frac1 2$) at base temperature in the superconducting parent compound Fe$_{1+x}$Te.   This critical concentration of interstitial iron corresponds to $x \approx 12\%$ and leads crystallographic phase separation at base temperature.  The spin-density wave is short-range ordered with a correlation length of 22(3) \AA, and as the ordering temperature is approached its propagation vector decreases linearly in the H-direction and becomes long-range ordered.  Upon further populating the interstitial iron site, the spin-density wave gives way to an incommensurate helical ordering with propagation vector $ \mathbf k =$ ($0.3855(2) \, \,0 \, \, \frac1 2$) at base temperature.  For a sample with $x \approx 9(1) \%$, we also find an incommensurate spin-density wave that competes with the bicollinear commensurate ordering close to the N\'eel point.  The shifting of spectral weight between competing magnetic orderings observed in several samples is supporting evidence for the phase separation being electronic in nature, and hence leads to crystallographic phase separation around the critical interstitial iron concentration of 12\%.  With results from both powder and single crystal samples, we construct a magnetic-crystallographic phase diagram of Fe$_{1+x}$Te for $ 5\% < x  <17\%$.  
\end{abstract}

\maketitle

\section{Introduction}

As in the high-$T_c$ cuprates, magnetism is implicated in the superconducting mechanism of the new Fe-based materials.  Detailed phase diagrams of  CeFeAsO$_{1-x}$F$_x$,\cite{Zhao_2008} and BaFe$_{2-x}$Co$_x$As$_2$,\cite{Nandi_2010} have revealed the proximity of a striped antiferromagnetic ordering to the superconducting regime.  Unsurprisingly, the parent phases of these superconductors are heavily studied to elucidate the possible role that magnetic ordering and crystal structure have on the electronic properties.  Structurally related to the iron pnictides but without the need for the compensating cationic layers, is the simple binary chalcogenide Fe$_{1+x}$Se, which was also found to be superconducting.\cite{Kotegawa_2008}   While isostructural to Fe$_{1+x}$Se, Fe$_{1+x}$Te does not exhibit bulk superconductivity unless there is sufficient anionic substitution of Te$^{2-}$ by either S$^{2-}$ or Se$^{2-}$.\cite{Yeh_2008, Fang_2008, Mizuguchi_2009}  The nonstoichiometry of Fe$_{1+x}$Te can be understood to arise from extra interstitial iron cations between the layers of edge-sharing FeTe$_4$ tetrahedra.  Here, we explore the crystal and magnetic structures of the parent phase Fe$_{1+x}$Te as a function of interstitial iron $x$ and temperature and evaluate the nature of its magnetic exchange interactions.

While anion substitution in Fe$_{1+x}$Te is isovalent, it does play a similar role to hole and electron doping in the FeAs-based materials as it suppresses a structural distortion so that the crystal structure remains tetragonal down to its ground state.  Interestingly, several studies on Fe$_{1+x}$Te have also revealed that there exists a correlation between the amount of anion substitution and the amount of interstitial iron, with the ``optimal doping'' of S$^{2-}$ or Se$^{2-}$ corresponding to a complete absence of interstitial iron.\cite{Sales_2009, Hu_2009, Zajdel_2010}   These two variables have been decoupled in two studies on the removal of interstitial iron topotactically through reaction of powder samples with iodine vapor.\cite{Rodriguez_2010, Rodriguez_2010b}  Indeed, the study on a series of Fe$_{1+x}$Te$_{0.7}$Se$_{0.3}$ powders without varying the Se/Te ratio demonstrated that the superconducting volume fraction was increased as $x$ was reduced to zero.\cite{Rodriguez_2010b}

\begin{figure*}
\includegraphics[width=1.0\linewidth,angle=0.0]{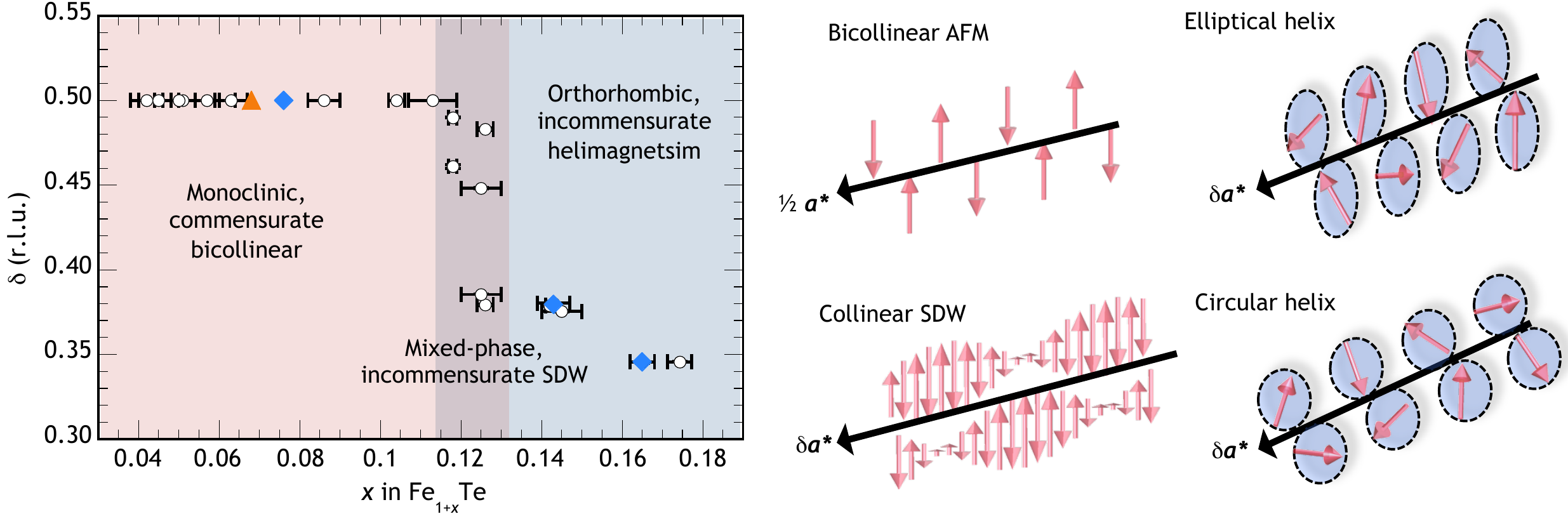}
\caption{[color online]  Magnetic-crystallographic phase diagram for Fe$_{1+x}$Te constructed by plotting the $\delta$ of the propagation vector \propinc versus concentration of interstitial iron at base temperature.  The open circles are for data from samples in this paper, triangle for data from Li \textit{et al.},\cite{S-Li_2009} and diamonds from Bao \textit{et al.}\cite{Bao_2009}  At right, the four different magnetic orderings in Fe$_{1+x}$Te observed in our neutron diffraction studies.  In the commensurate bicollinear antiferromagnetic (AFM) phase, the moments are along the $b$-direction only.  Upon increasing interstitial iron to 12\%, the bicollinear AFM phase gives way to an incommensurate spin-density wave (SDW) that is collinear and with moments pointed along the $b$-direction.   Upon further increasing $x$, a spin component develops along the $c$-direction, creating first an elliptical helix (elongated along $b$-direction) and then circular helix phase.  The direction of the propagation vector is shown for all.}
\label{FeTe_phases}
\end{figure*}

One way the iron chalcogenides differ remarkably from the FeAs-based superconductors is in the nature of the antiferromagnetic ordering. In the FeAs-based systems, the magnetic structure is described as a collinear striped ordering, which is termed ($\pi$, $\pi$) ordering since it corresponds to a wave vector connecting the $\Gamma$ to $M$ points in the Brillouin zone.  Contrastingly, in Fe$_{1+x}$Te the long-range magnetic ordering is a bicollinear structure that is rotated by $45^{\circ}$ with respect to the ordering of the iron arsenides.\cite{S-Li_2009}    This structure corresponds to a wave vector of ($\pi$, 0), but can change to incommensurate ($\delta \pi$, 0) with greater amounts of interstitial iron.\cite{Bao_2009}   Furthermore, the size of the magnetic moment per Fe cation in Fe$_{1+x}$Te is $\approx$ 2 $\mu_B$, much larger than those in analogous parent phases, \textit{e.g.} 0.36(5) $\mu_B$ in LaOFeAs,\cite{delaCruz_2008} 0.93(6) $\mu_B$ in BaFe$_2$As$_2$,\cite{Wilson_2009} and 0.09(4) $\mu_B$ in NaFeAs.\cite{S-Li_2009}   The magnetic properties of the arsenides have led several researchers to describe the observed ordering to be due to nesting of the Fermi Surface and therefore largely due to itinerant electron behavior.\cite{Mazin_2008b}  For Fe$_{1+x}$Te, experimental evidence points to a localized model with magnetic susceptibility measurements showing that it follows Curie-Weiss behavior.\cite{Chiba_1955, S-Li_2009}

The differences in magnetic ordering raise the possibility that a local moment picture best describes the magnetism in Fe$_{1+x}$Te and that the chalcogenides are fundamentally different types of superconductors from the FeAs-based ones.  Despite the discrepancies outlined above, there are some conspicuous similarities between the two systems within the superconducting state; for example, the so-called spin resonance has been observed as a gapped excitation in inelastic neutron scattering experiments.  The spin resonance is located at the ($\pi$, $\pi$) position, with an energy that scales with $T_c$.  In the Ba$_{1-x}$K$_x$Fe$_2$As$_2$ it was found to be $\approx 14$ meV,\cite{Christianson_2008} and in Fe$_{1+x}$Te$_{0.6}$Se$_{0.4}$ observed around 7 meV.\cite{Qiu_2009}  Thus, several studies have focused on the peculiar change of the magnetic ordering wavevector from ($\pi$, 0) to ($\pi$, $\pi$) in Fe$_{1+x}$Te$_{1-y}$Se$_y$ as a function of Se substitutution.\cite{Babkevich_2010, Xu_2010, Liu_2010}  A central question concerning iron telluride is whether the effect from anionic substitution is either to suppress a structural distortion, or just remove interstitial iron from the lattice.  Of course, another possibility is that both are necessary for superconductivity.  Comprehensive reviews on iron-based superconductors and more particularly the iron chalcogenides can be found elsewhere.\cite{Johnston_2010, Paglione_2010, Lynn_2009}

Here we study the parent phase Fe$_{1+x}$Te for different amounts of $x$ to understand how chemical composition controls the crystal structure and magnetic ordering.  We have prepared several single crystal and powder samples and have outlined key structural and magnetic parameters of the parent compound as a function of temperature and interstitial iron.  This led us to construct a phase diagram of Fe$_{1+x}$Te at base temperature (5 K to 15 K) for varying amounts of $x$ (Fig.~\ref{FeTe_phases}).   The resulting phase diagram for Fe$_{1+x}$Te shows that the magnetic ordering is richer than found previously and that the propagation vector and crystal structure undergo an abrupt change at a critical amount of interstitial iron, $x \approx 12\%$.  Throughout this paper we describe how studies with neutron powder diffraction in combination with polarized neutron single crystal diffraction has allowed us to determine the relationship between the crystal structure and magnetic ordering for samples of Fe$_{1+x}$Te for $ 5\% < x  <17\%$.  The polarized neutron studies distinguish between structures that are spin amplitude modulated (\textit{i.e.} spin density wave) versus those that are spin direction modulated (\textit{i.e.} helical or cycloidal ordering), which were all found in this system (Fig.~\ref{FeTe_phases}).  The results are divided according to the different types of ordering observed for $x < 12\%$, $x > 12\%$, and $x \approx 12\%$.   We then discuss possible exchange couplings including both the interstitial iron and in-plane iron to explain the diverse orderings observed in Fe$_{1+x}$Te.

\section{Experiments}

The powder samples were prepared by combining nominal amounts of iron and tellurium powders in evacuated quartz ampoules, after grinding them in a mortar and pestle.  The powder mixtures were first heated to 450$^{\circ}$ C for a soak time of 12 hrs, followed by a slow ramp up to 750$^{\circ}$ C for 12 hours, after which they were furnace cooled.  The single crystal samples were prepared by heating pre-made powder samples up to 820$^{\circ}$  C to 850$^{\circ}$  C under vacuum, with the higher melt temperature for samples with higher iron concentrations.  In past studies of the Fe-Te phase diagram,\cite{Ipser_1974, Mann_1977, Saha_1988, Okamoto_1990} the melting point seems to vary upon iron concentration but the maximum is around 844 to 847$^{\circ}$ C at standard temperature and pressure.   The samples were kept above the melting point for 12 hours and then slow-cooled at a rate of 6$^{\circ}$ C/hr.  

The amount of interstitial iron, as determined by diffraction measurements, corresponded to approximately 1 \% to 3 \% less iron than the nominal amount, presumably due to reaction of the iron with the quartz ampoule or from oxide contamination in the starting iron powder, as previously reported for the preparation of superconducting Fe$_{1+x}$Se samples.\cite{McQueen_2009b} For the samples with high iron concentration, this meant adding additional iron powder to the mixture in the second reheat and crystal growth process.  The amount of interstitial iron that can be accommodated in the layered $\beta$-phase has been reported by several studies, and can range from 7.5\% to 16.3\%,\cite{Mann_1977} or a much narrower 4.2\% to 8.7\%.\cite{Ipser_1974, Okamoto_1990}  Above the maximum amount, the $\beta$-phase is shown to be in equilibrium with iron metal.  From our single X-ray and neutron powder diffraction studies presented below, the amount of interstitial iron in our samples vary between 4.2(4)\% to 17.4(4)\%.

Since these layered compounds cleave easily along the (0 0 1) plane, small crystals were cleaved from the larger crystals for single crystal X-ray diffraction in order to characterize the amount of interstitial iron.  The single crystal XRD was performed with Mo K$\alpha$ radiation ($\lambda = 0.71073 \AA$) and the data collected at 250 K.

The powder samples were characterized by the BT-1 diffractometer at the NIST Center for Neutron Research (NCNR) with a wavelength of $\lambda = 2.0785$ \AA~(Ge311 monochromator).  Spallation source neutron diffraction was also performed on select samples using the NPDF powder diffractometer at the Lujan Neutron Center at the Los Alamos National Laboratory.

Single crystal neutron diffraction experiments were performed on several spectrometers at the NCNR.  The characterization of the propagation vector for several crystals was performed on the BT-9 thermal triple axis spectrometer with a $\lambda = 2.0875$ \AA~(pyrolytic graphite monochromator).  Two-dimensional maps of the (H 0 L) plane of a 200 mg single crystal were also obtained on the MACS cold-source spectrometer.  In MACS, only elastic scattering planes were scanned by fixing the final and incident energies to 3.6 meV using the 20 double bounce PG(002) analyzing crystals and a double focused PG(002) monochromator.

\begin{figure}[!t] \centering
\includegraphics[width=1.0\linewidth,angle=0.0]{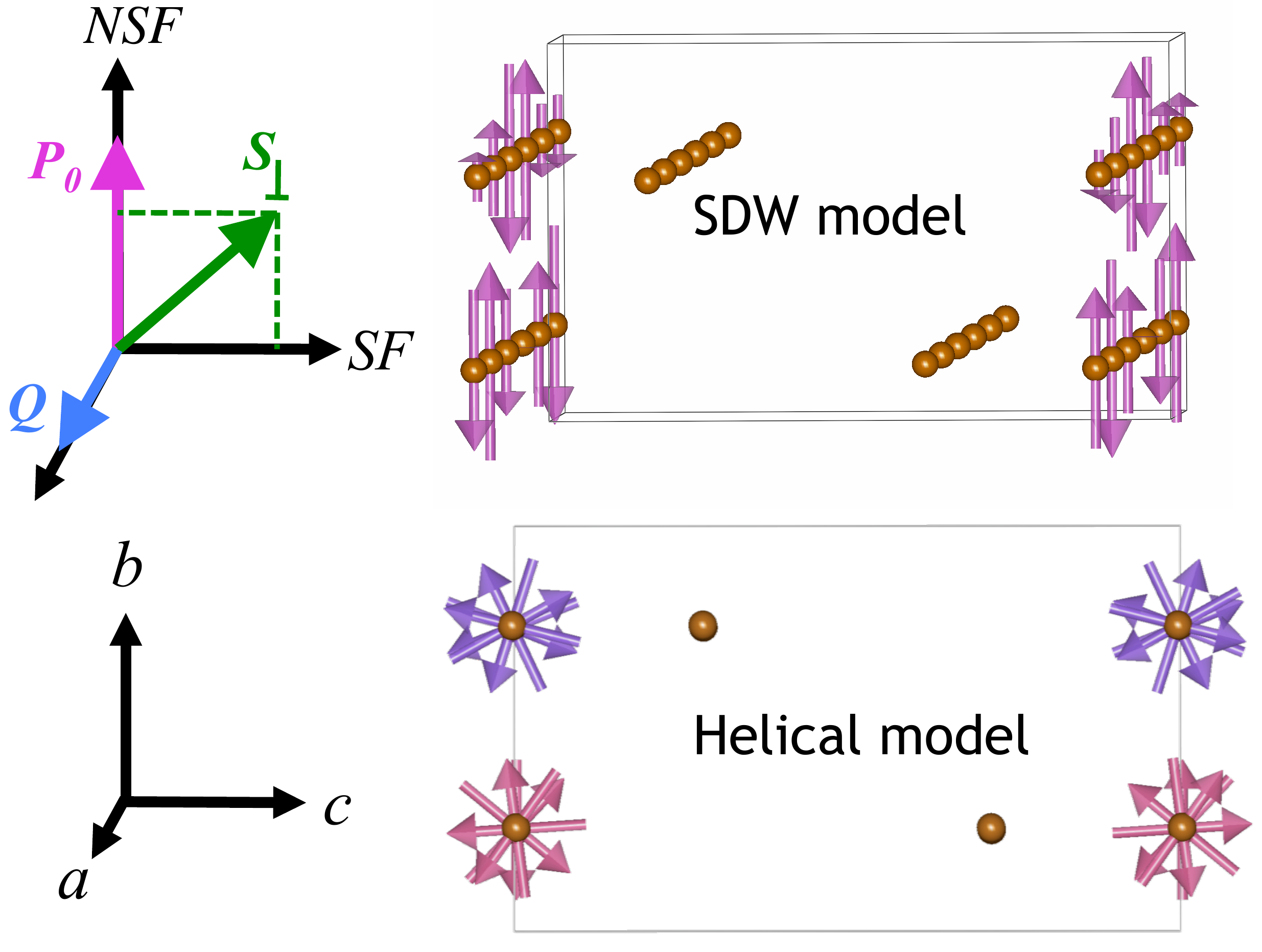}
\caption{[color online]  The experimental setup for the neutron spin polarized diffraction studies on single crystals of Fe$_{1+x}$Te.  The neutron beam is polarized vertically as represented by the vector $\mathbf P_0$, which is normal to the scattering vector $\mathbf Q$, allowing one to measure the spin amplitude vector $\mathbf{S}_{\perp}$ in the non-spin flip (NSF) and spin flip (SF) channels.  By aligning the crystals to have the $b$-axis parallel to $\mathbf P_0$,  one can distingish between a magnetic structure with collinear arrangement as the spin density wave (SDW) model or a non-collinear one such as the helical model.  The interstitial iron site is shown in the crystal structures, but their moments (along with the Te atoms) are excluded for clarity.}
\label{FeTe_SPINS}
\end{figure}

Polarized neutron diffraction was performed on a 300 mg Fe$_{1.09(1)}$Te single crystal and the same crystal used in the MACS experiment.  The measurements were performed on the SPINS cold neutron spectrometer with $\lambda = 4.0449$ \AA~and a beam polarized vertically using supermirrors.  The thin Fe/Si magnetic films within the supermirror reflect spin $+\frac1 2$ neutrons, so only spin $-\frac1 2$ neutrons are transmitted, the latter of which were incident on the sample.  Polarization analysis of the reflected beam was performed with a similar Soller collimator and supermirror assembly described in earlier work.\cite{majkrzak_1995, Rodriguez_2011}  Tight collimation following the supermirrors was used to absorb the $+\frac1 2$ neutrons, and flipper coils were placed before and after the sample.

The crystals were aligned so the scattering vector $\mathbf{Q}$ was set perpendicular to the beam polarization direction  $\mathbf{P}_0$, and $\mathbf{S}_{\perp}$ was measured.   We define $\mathbf{S}_{\perp}$  as the spin amplitude vector normal to $\mathbf{Q}$.  In the non-spin flip (NSF) channels, the $\mathbf{S}_{\perp}$ component parallel to the $b$-axis is measured, and in the spin flip (SF) channels, component parallel to the (H 0 L) plane is measured.  This experimental setup is illustrated in Fig.~\ref{FeTe_SPINS}.  Nuclear scattering also appears in the NSF channels and the (001) nuclear peak was measured for both crystals to obtain the NSF/SF ratio, or flipping ratio, which was found to be $\approx 20$.

\section{Results}

\subsection{Crystallography}

\begin{figure}[b] \centering
\includegraphics[width=0.9\linewidth,angle=0.0]{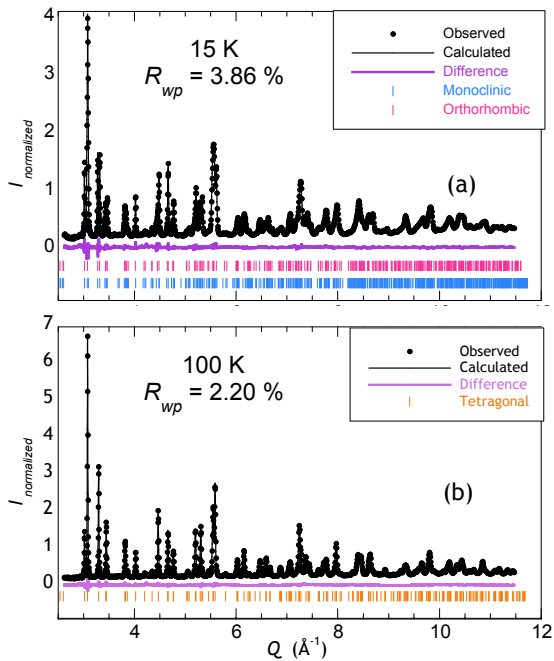}
\caption{[color online]  The observed and calculated neutron powder diffraction patterns of Fe$_{1.12}$Te from the time-of-flight NPDF diffractometer with the difference pattern and phase reflection marks below.  In (a) the 15 K data is fit with monoclinic and orthorhombic phases, and in (b) the 100 K data is fit with a single tetragonal phase.}
\label{FeTe_NPD}
\end{figure}

Below the the N\'eel point  ($\approx$ 60 K to 70 K), Fe$_{1+x}$Te is known to undergo a crystallographic phase transition from tetragonal $P4/nmm$ symmetry to either monoclinic $P2_1/m$ or orthorhombic $Pmmn$ symmetry.\cite{Fruchart_1975, S-Li_2009, Bao_2009}  Therefore, the neutron powder diffraction (NPD) studies on 13 samples were performed at 100 K and base temperature (5 K to 15 K).    The structural parameters for all the samples were obtained using the GSAS Rietveld suite of programs.\cite{GSAS}  Although no new crystallographic phases were found, there is a special iron concentration of $x\approx 12\%$ that leads to phase separation at base temperature as shown in Fig.~\ref{FeTe_NPD}a.   At 100 K this phase can be fit with a single tetragonal phase (Fig.~\ref{FeTe_NPD}b) even with data from the high-resolution, backscattering banks of NPDF and the high-resolution BT-1 diffractometer.  The structural parameters for this phase at base temperature and 100 K from the NPDF data are given in Table~\ref{crystal_data}.  The structural parameters for the phase with a higher amount of interstitial iron Fe$_{1.142(1)}$Te are also presented in Table~\ref{crystal_data}.

\begin{table}[b]
\caption{Crystal structural parameters for Fe$_{1.119(1)}$Te and Fe$_{1.142(1)}$Te powder samples obtained from the NPDF data at 15 K and 100 K.}
\begin{tabular}{c c l l p{0.5cm} l l l}
\hline
\hline
\multicolumn{8}{c} {Fe$_{1.119(1)}$Te, 15 K,  $R_{wp} = 3.86\%$ } \\
\hline
\multicolumn{8}{c} { $P2_1/m$ (unique axis $b$)} \\
\multicolumn{8}{c} {$a = 3.83378$(6), $b = 3.78667$(8), $c = 6.246427$(8), $\beta = 89.359$(1)}\\
atom	&        Site  & 	               x	 &         y   & {} &	                    z	& $U_{iso}$ (\AA$^2$)  &	Occ. \\
Fe1	&          2e	 &	0.7600(3)  &  	0.25 & {} &	0.0036(2)  &  		0.00206(2) & 	1.0  \\
Fe2	&          2e	 &	  0.240(2)  &      0.25 & {} &	   0.715(1) &  		0.00206(2) & 0.119(1)  \\
Te	&          2e	 &	0.2552(4)  &      0.25 & {}	&        0.2844(2) &  		0.00206(2) & 	1.0  \\
\\
\multicolumn{8}{c} {$Pmmn$ (origin choice 2)} \\
\multicolumn{8}{c} {$a = 3.83259$(5), $b = 3.78667$(8), $c = 6.24627$(8)} \\
atom	&        Site  & 	            x  & 	y	 & {} &	          z	&$U_{iso}$ (\AA$^2$) &	    Occ. \\
Fe1	&          2b	 & 	      0.75  & 	0.25 & {} &	0.0028(3) &  		  0.00206(2)&	 	1.0  \\
Fe2	&          2a	 & 	      0.25  & 	0.25 & {} &           0.726(2) &  		  0.00206(2)& 0.119(1)   \\
Te	&          2a	 & 	      0.25  &		 0.25 & {} &	0.2821(3) &  		  0.00206(2)&	 	1.0  \\
\\
\hline
\multicolumn{8}{c} {Fe$_{1.0.119(1)}$Te, 100 K, $R_{wp} = 2.20\%$  } \\
\hline
\multicolumn{8}{c} {$P4/nmm$ (origin choice 2)} \\
\multicolumn{8}{c} {$a = 3.8119$(1), $c = 6.2468$(2)} \\
atom	&        Site  & 	            x  & 	y	 & {} &	          z	& $U_{iso}$ \AA$^2$ &	  Occ. \\
Fe1	&          2a	 & 	      0.75  & 	0.25 & {} &	            0.0 & 		 0.00528(2)& 	1.0  \\
Fe2	&          2c	 & 	      0.25  & 	0.25 & {} &         0.7220(3) &  		 0.00528(2) &   0.119(1) \\
Te	&          2c	 & 	      0.25  &		 0.25 & {} &      0.28367(7) &  		 0.00528(2) &    1.0  \\
\\
\hline
\multicolumn{8}{c} {Fe$_{1.142(1)}$Te, 15 K, ,  $R_{wp} = 3.42\%$  } \\
\hline
\multicolumn{8}{c} {$Pmmn$ (origin choice 2)}\\
\multicolumn{8}{c} {$a = 3.81856$(4), $b = 3.79092$(4), $c = 6.24898$(7)} \\
atom	&        Site  & 	            x  & 	y	 & {} &	          z	&$U_{iso}$ (\AA$^2$) &	    Occ. \\
Fe1	&          2b	 & 	      0.75  & 	0.25 & {} &	0.0029(2) &  		  0.00353(5)&	 	1.0  \\
Fe2	&          2a	 & 	      0.25  & 	0.25 & {} &         0.6954(6) &  		  0.00353(5)& 0.142(1)   \\
Te	&          2a	 & 	      0.25  &		 0.25 & {} &	0.2801(1) &  		  0.00353(5)&	 	1.0  \\
\\
\hline
\multicolumn{8}{c} {Fe$_{1.142(1)}$Te, 100 K, $R_{wp} = 3.72\%$  } \\
\hline
\multicolumn{8}{c} {$P4/nmm$ (origin choice 2)}\\
\multicolumn{8}{c} {$a = 3.81141$(2), $c = 6.24656$(7)} \\
atom	&        Site  & 	            x  & 	y	 & {} &	          z	& $U_{iso}$ \AA$^2$ &	  Occ. \\
Fe1	&          2a	 & 	      0.75  & 	0.25 & {} &	            0.0 & 		 0.00506(2)& 	1.0  \\
Fe2	&          2c	 & 	      0.25  & 	0.25 & {} &         0.6953(6) &  		 0.00506(2) &   0.142(1) \\
Te	&          2c	 & 	      0.25  &		 0.25 & {} &      0.2801(1) &  		 0.00506(1) &    1.0  \\
\hline
\hline
\end{tabular}
\label{crystal_data}
\end{table}

The neutron powder diffraction results confirm that increasing interstitial iron changes the low-temperature phase from monoclinic to orthorhombic symmetry.  We find that 11.9(1)\% of excess iron is the percentage necessary to nucleate the orthorhombic phase.

\begin{figure}[!t] \centering
\includegraphics[width=1.0\linewidth,angle=0.0]{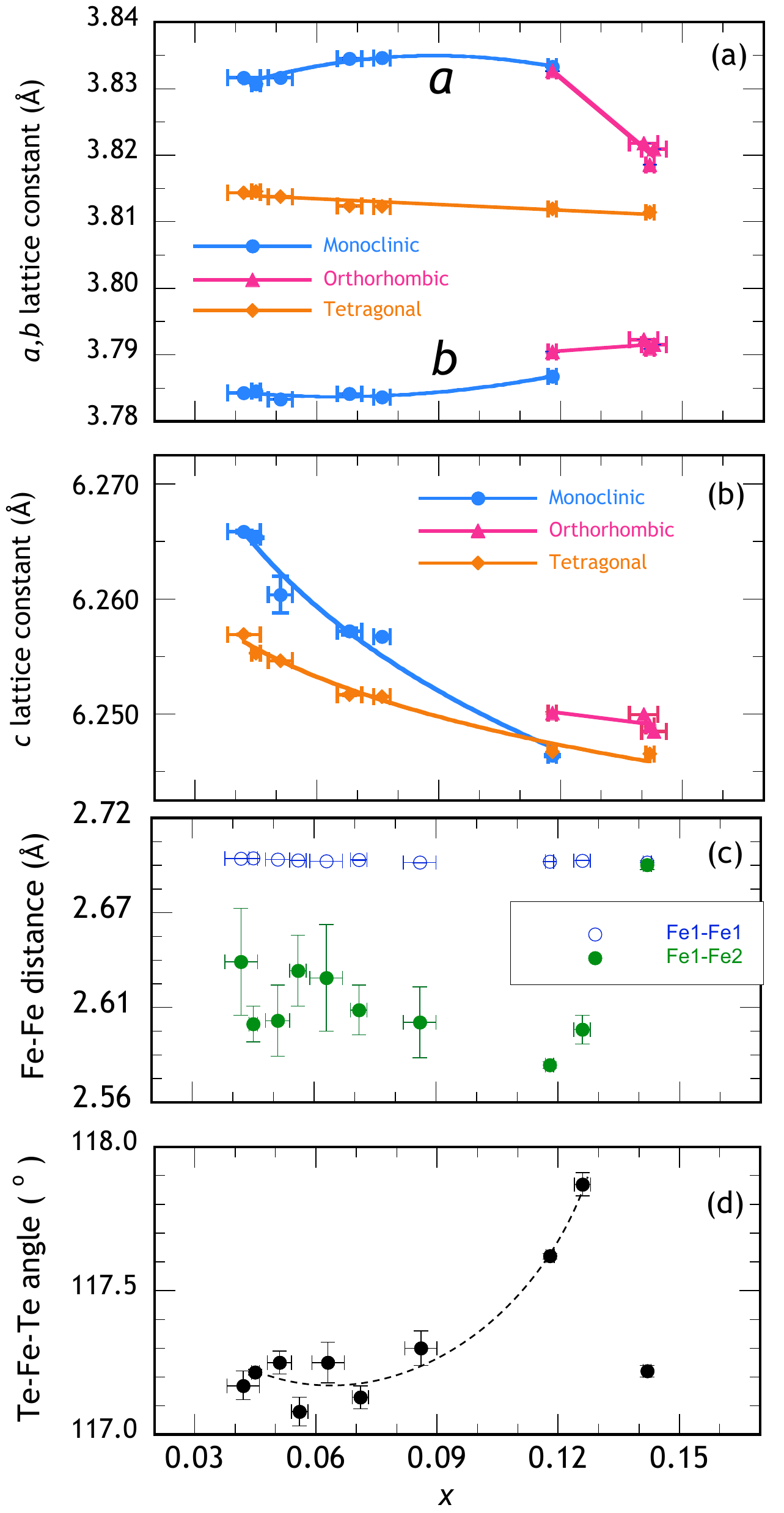}
\caption{[color online]  Lattice parameters and relevant bond distances and angles as a function of interstitial iron, $x$, obtained from the neutron analysis.  (a) The $a,b$ lattice constants with the tetragonal data taken at 100K, and the rest at base temperature (5 K to 15 K).  (b) The $c$ lattice constant, or interplanar spacing. (c) The iron-iron bond distances with Fe1 corresponding to the in-plane iron and Fe2 to the interstitial iron at 100 K.  (d)  The Te--Fe1--Te tetrahedral bond angle at 100 K.}
\label{FeTe_params}
\end{figure}

Lattice constants and relevant bond distances and angles from the BT-1 and NPDF data are presented in Fig.~\ref{FeTe_params}.  The splitting of the $a$-parameter at the low temperature transition is quite dramatic but remains mostly constant throughout the monoclinic phase (Fig.~\ref{FeTe_params}a).  In the orthorhombic phase, the splitting between the $a$ and $b$ parameters is reduced.  The $c$-parameter, which corresponds to interlayer spacing, decreases as the amount of interstitial iron is increased (Fig.~\ref{FeTe_NPD}a,b).  This trend makes sense since the coordination of the interstitial iron to the Te anions is square pyramidal, bonded to four Te atoms within one layer and a fifth one in the adjacent layer.  As more of these interstitial sites are occupied, the effect should be to draw the layers together and therefore decrease $c$.  This trend was found for all the crystallographic phases (Fig.~\ref{FeTe_params}b).

The small percent occupancy and disordered nature of the interstitial iron site significantly increase the standard uncertainties of its structural parameters.  Furthermore, the fractional coordinates are correlated to atomic displacement parameters $U_{iso}$'s and occupancies.  In the refinements, the $U_{iso}$'s were constrained to be equal for all the atoms.  Since only one coordinate is refinable for the interstitial iron (Fe2 in Table~\ref{crystal_data}) in the tetragonal phase, the relevant bond distances and bond angles presented in Fig.~\ref{FeTe_params}c,d are from the tetragonal phase.   The Fe1-Fe1 distance is the nearest neighbor distance within the iron square lattice (= $a/\sqrt 2$) and Fe1-Fe2 is the distance from the in-plane iron site (Fe1) to the interstitial iron site (Fe2).  The iron-iron distances are shown in Fig.~\ref{FeTe_params}c, and it is remarkable that for most of the phase diagram, the Fe1-Fe1 distance is smaller than the Fe1-Fe2 distance.     Only when excess iron reaches $\approx 14\%$ does the Fe1-Fe2 distance become equal to that of Fe1-Fe1( Fig.~\ref{FeTe_NPD}c).  No doubt this shift in iron-iron distances causes a change in exchange parameters that would explain the different magnetic structures due to varying amounts of interstitial iron.

Another interesting parameter to observe upon increasing $x$ in Fe$_{1+x}$Te, is the Te--Fe--Te tetrahedral bond angle.   This tetrahedral angle along with pnictide/chalcogenide height have been cited as important structural parameters in the Fe-based superconductors.  Generally, the closer this angle gets to the ideal 109.5 $^{\circ}$, the higher the $T_c$.\cite{Zhao_2008, Lee_2008}  As interstitial iron is increased, this angle becomes more distorted in the monoclinic phase until critical percentage of $\approx12\%$ above which the structure becomes orthorhombic (Fig.~\ref{FeTe_params}d).   Within the monoclinic phase, the $a$ and $b$ parameters are not changing significantly with interestitial iron, unlike the interlayer spacing and the Te--Fe--Te bond angles.  Therefore, the structural change from monoclinic to orthorhombic symmetry could be driven by the lattice lowering its energy by retaining the Te--Fe--Te bond angle closer to $\approx 117.2^{\circ}$, a value common to both the low and high end of $x$ in Fe$_{1+x}$Te (Fig.~\ref{FeTe_params}d).

\subsection{Collinear magnetic ordering for $x< 12\%$}

\begin{figure}[!b] \centering
\includegraphics[width=1.0\linewidth,angle=0.0]{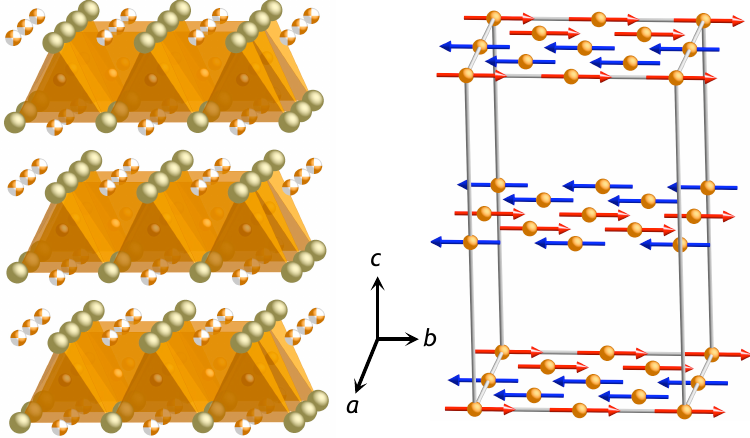}
\caption{[color online]  The crystal structure of Fe$_{1+x}$Te with the layers consisting of edge-sharing FeTe$_4$ tetrahedra.  The interstitial iron sites, shown as beach ball spheres, are partially occupied and disordered.  The magnetic lattice of the antiferromagnetic structure commensurate with the chemical lattice consists of bicollinear chains with moments pointing in the $b$-direction.  Only the moments of the Fe atoms in the tetrahedral coordination are shown for clarity. Such ordering corresponds to a magnetic propagation vector of \propcom in reciprocal lattice units.}
\label{FeTe_structure}
\end{figure}

The magnetic Bragg peaks observed in the BT-1 powder data of Fe$_{1.051(3)}$Te can be fit with the commensurate magnetic structure known as a bicollinear antiferromagnetic structure (Fig.~\ref{FeTe_structure}).  Although this is a straightforward collinear ordering, the method of respresentational analysis was employed here in order to be consistent with the analysis of the more complex, incommensurate orderings presented for $x >12$.  Representational analysis using the program BasIreps (version 4.0) from the FullProf Rietveld suite was employed,\cite{fp} and the irreducible representations with their basis vectors for vector \propcom under $P2_1m$ symmetry are presented in Table~\ref{irreps}.

\begin{table}[!]
\caption{The basis functions $\psi$ for each Fe atom in the unit cell under the four irreducible representations for both space groups $P2_1m$ and $Pmmn$.  The return vector $\epsilon$ is $exp(-i \pi\delta)$, where $\delta$ is part of the propagation vector \propinc and varies according to amount of interstitial iron $x$.  The coordinates for site 1 are  $x$,$y$,$z$, and those for site 2 are -$x$,$y+\frac{1}{2}$,-$z$ for $P2_1m$ and $x+\frac{1}{2}$,-$y$,-$z$ for $Pmmn$.}
\begin{tabular}{c|cc|cc}
\hline
\hline
{}                       & \multicolumn{2}{c}{$P2_1/m$} &  \multicolumn{2}{c}{$Pmmn$}  \\
Irrep 		        &    $\psi$ for site 1    & $\psi$ for site 2    &    $\psi$ for site 1         & $\psi$ for site 2    \\
\hline
$\Gamma_1$  & (0 1 0)                 & (0 1 0)    & (0 1 0)          & (0 $\epsilon$ 0)    \\
$\Gamma_2$  & (1 0 0)                 & (-1 0  0)  & (1  0  0)        & (-$\epsilon$  0 0)  \\
{}                      & (0 0 1)                 & (0 0 -1)   & (0  0 1)         & (0 0 $\epsilon$)    \\
$\Gamma_3$  & (1 0 0)                 & (1 0 0)    & (1 0 0)          & ($\epsilon$  0  0)  \\
{}                      & (0 0 1)                 & (0 0 1)   & (0  0  1)         & (0 0 -$\epsilon$)   \\
$\Gamma_4$ & (0 1 0)                  & (0 -1 0)   &  (0 1 0)           & (0 -$\epsilon$ 0)  \\
\hline
\hline
\end{tabular}
\label{irreps}
\end{table}

\begin{figure}[!b] \centering
\includegraphics[width=1.0\linewidth,angle=0.0]{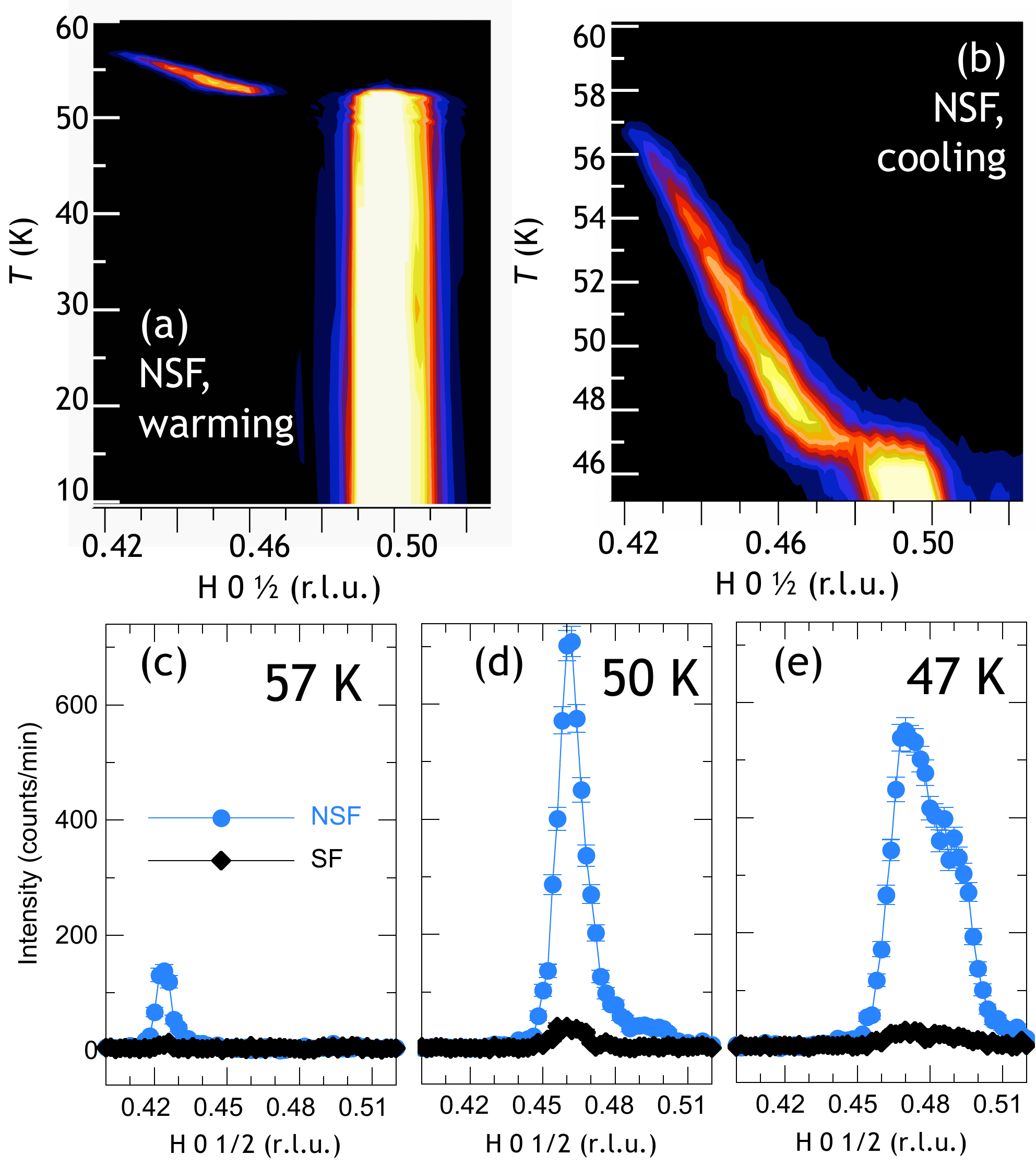}
\caption{[color online]  The non-spin flip (NSF) magnetic scattering of a single crystal with composition Fe$_{1.09(1)}$Te as measured on the SPINS spectrometer with a vertically polarized neutron beam.  In (a) the contour map shows the scattering versus temperature upon warming, and in (b) upon cooling.    Near the N\'eel point, an  incommensurate wave vector is shown in both maps to compete with the commensurate \propcom ordering.  In (c) through (e), cross sections of the scattering in the NSF and spin-flip (SF) channels is shown for various temperatures upon cooling.  This data confirms a model where the moment lies only in the $b$-direction for both the incommensurate wave vector appearing close to the N\'eel point and the commensurate wave vector that exists in the ground state of Fe$_{1.09(1)}$Te}.
\label{FeTe_comm_SPINS}
\end{figure}

The magnetic Bragg peaks were fit with representation $\Gamma_1$, which shows that the moment has a component only in the $b$-direction and that the iron atoms at $x$,$y$,$z$ and -$x$,$y+\frac{1}{2}$,-$z$ are ferromagnetically aligned.  This leads to the bicollinear ordering, which features two ferromagnetically coupled stripes (Fig.~\ref{FeTe_structure}).  The average moment size obtained from three powder samples is 1.78(3) $\mu_B$/Fe for both the in-plane and interstitial iron sites.  This value is close to the one reported by Ikkubo \textit{et al.} of 1.86(2) $\mu_B$\cite{Ikkubo_2009} but lower than the values of 2.54(2), 2.25(8), and 2.07(7) $\mu_B$ found by previous neutron studies.\cite{Martinelli_2010, S-Li_2009, Fruchart_1975}  One possible reason for this range in reported moment size could be due to some studies allowing the moment to point in any direction,\cite{S-Li_2009, Fruchart_1975}  while the 1.78(3) $\mu_B$ found in this study is obtained when the moment is constrained to be along $b$.  The magnetic structure of all our powder samples with interstitial iron less than 12\% were successfully fit with this representation.

Polarized neutron measurements on a single crystal sample (SPINS) confirm that the collinear structure with moments only along the $b$-axis is the correct model for the commensurate phase.  If scattering is be observed only in the NSF channels, the ordering is collinear with moments only along $b$ ($\Gamma_1$ representation).  If there is also scattering in the SF channels, then helical or other noncollinear ordering is correct  (Fig.~\ref{FeTe_SPINS}).  For a crystal with composition Fe$_{1.09(1)}$Te, only scattering in the NSF channel was observed (Fig.~\ref{FeTe_comm_SPINS}), which would be consistent with the model of spin component only along $b$.  

Contour maps of the NSF magnetic scattering versus temperature upon warming and cooling are shown in Fig.~\ref{FeTe_comm_SPINS}a,b.   One interesting feature not observed before appears close to the N\'eel point; an incommensurate propagation vector competes with the commensurate bicollinear ordering.  Upon cooling and warming, this incommensurate wave vector appears at H$=0.421(1)$ and moves towards the commensurate H = 0.5 position (Fig.~\ref{FeTe_comm_SPINS}b).  It is important to note that this incommensurate wave vector, like the commensurate one, has no scattering in the SF channels (Fig.~\ref{FeTe_comm_SPINS}c--e) proving that the moment direction in this composition is collinear for both the commensurate and incommensurate orderings.

\begin{figure}[!t] \centering
\includegraphics[width=1.0\linewidth,angle=0.0]{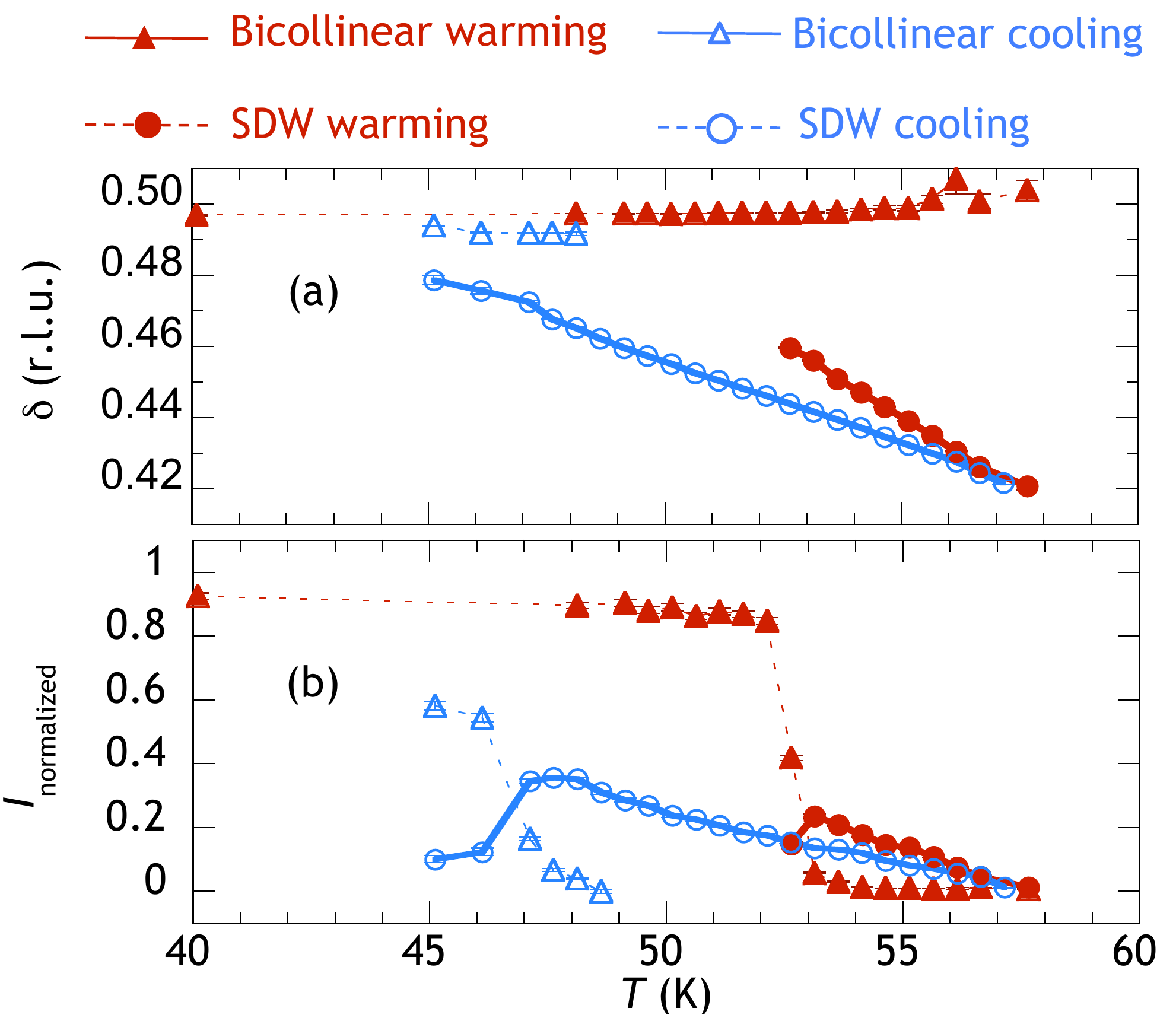}
\caption{[color online]  Peak centers and integrated intensities from fits to the magnetic Bragg peaks of a single crystal Fe$_{1.09(1)}$Te measured in the spin polarized experiments on the SPINS spectrometer.  In (a) the peak centers and therefore the $\delta$ from the propagation vector \propinc is shown versus temperature upon warming and cooling.  In (b) the integrated intensities for the Fe$_{1.09(1)}$Te crystal upon cooling and warming.}
\label{FeTe_SPINS_analysis1}
\end{figure}

The collinear incommensurate scattering seen close to the N\'eel point has never been observed for the composition known to have a bicollinear ordering as its ground state.  The competition between the two types of ordering is better presented by plotting the peak centers, integrated intensities, and widths from Gaussian fits.  In Fig.~\ref{FeTe_SPINS_analysis1}a the peak center is plotted versus temperature upon warming and cooling; the incommensurate vector is shown to have a temperature dependence that is linear while the commensurate vector has little temperature dependence.  In Fig.~\ref{FeTe_SPINS_analysis1}b, the integrated intensities for both magnetic peaks at \propcom and \propinc are shown normalized to the magnetic intensity of \propcom at 10 K.  Upon warming, the spectral weight of the commensurate ordering shifts to the incommensurate one before the N\'eel point.  The hysteresis in the temperature dependence of the propagation vectors and integrated intensities suggests a first-order transition, which is consistent with a structural transition above the magnetic ordering one.

\subsection{Helical magnetism in $x>12\%$}

The incommensurate magnetic ordering of a powder sample (BT-1) with composition Fe$_{1.143(3)}$Te was solved with representational analysis since use of colored space groups or Shubnikov groups is insufficient to solve such structures.\cite{bertaut_1962, bertaut_1968}  For the crystal symmetry of $Pmmn$, there are four symmetry elements under which the propagation vector \propinc remains unchanged, leading to the four irreducible representations; their basis vectors $\psi$ are presented in Table~\ref{irreps}.  For the incommensurate structure of Fe$_{1.143(3)}$Te, the refinements using $\Gamma_1$, $\Gamma_2$, and $\Gamma_3$ converge, but only $\Gamma_1$ fits the powder profile satisfactorily (magnetic $R$ factor = 33.3\%).  This representation is related to that of $\Gamma_1$ under $P2_1/m$ symmetry (Table \ref{irreps}) since it leads to ferromagnetic coupling between the same atoms with the chief difference being that in $Pmmn$ the coupling is modulated by the phase factor $\pi \delta$.  The resulting structure from the fit to $\Gamma_1$ leads to a spin-amplitude modulated structure or spin-density wave (SDW) ordering illustrated in Fig.~\ref{FeTe_SPINS}.

\begin{figure}[!b] \centering
\includegraphics[width=0.9\linewidth,angle=0.0]{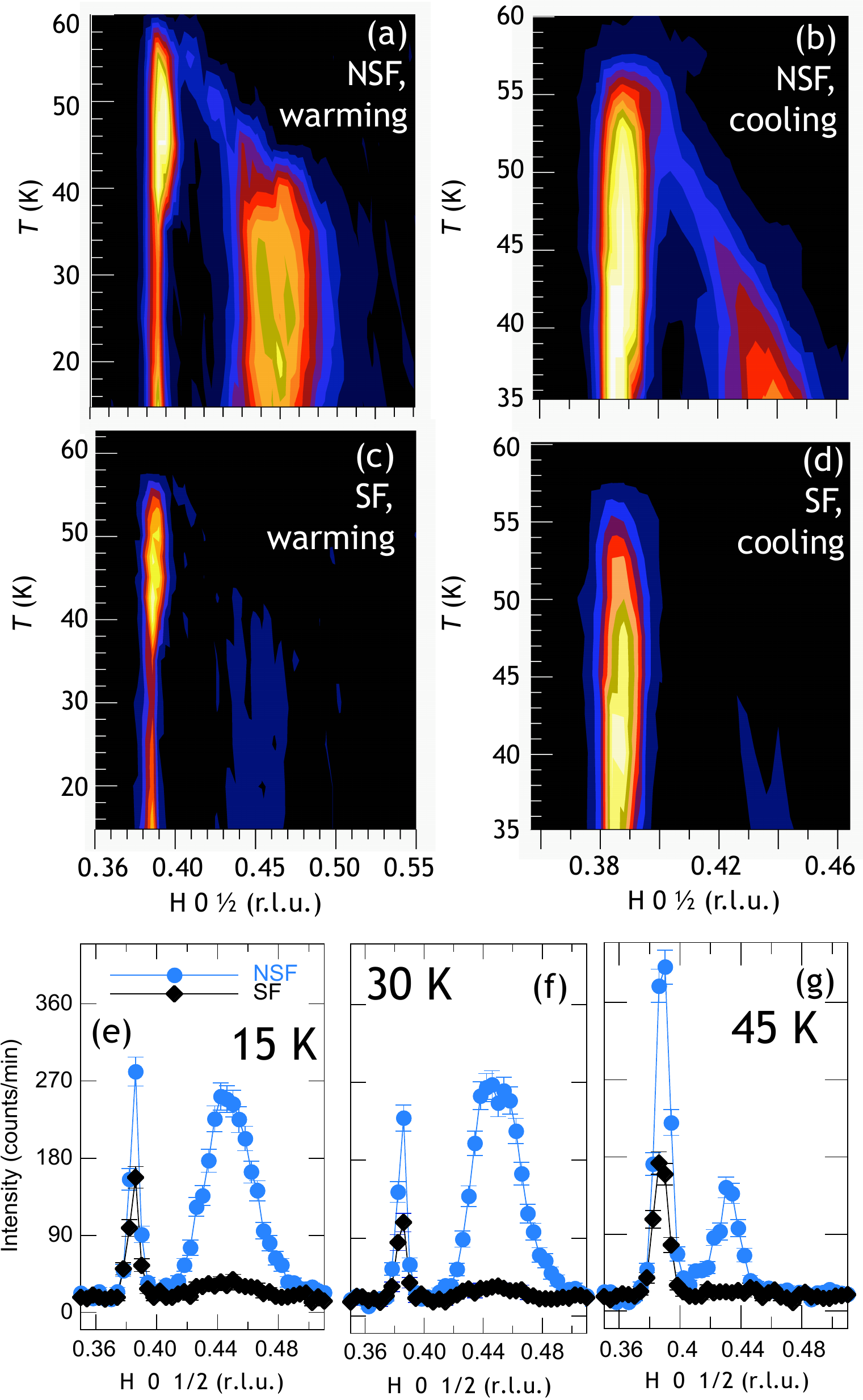}
\caption{[color online]  The non-spin flip (NSF) and spin flip (SF) magnetic scattering of a single crystal with composition Fe$_{1.124(5)}$Te as measured on the SPINS spectrometer with a vertically polarized neutron beam.  In (a) the contour map shows the NSF scattering versus temperature upon warming, and in (b) upon cooling.     In (c) the contour map shows the SF scattering versus temperature upon warming, and in (d) upon cooling.   What both measurements reveal is that the resolution-limited magnetic peak at H = 0.3855(2) corresponds to a helical type of ordering whereas the broad magnetic peak at H = 0.4481(4) corresponds to a spin density wave (SDW).  In (e) through (g), cross sections of the scattering in the NSF and SF channels is shown for various temperatures upon warming.}
\label{FeTe_1pER_SPINS}
\end{figure}

The study by Bao \textit{et al.}, however, found that the incommensurate structure is a complex helical structure with a moment contribution in all directions.  Since Fe$_{1+x}$Te undergoes a first order transition, the mixing of irreducible representations is allowed.  All combinations of the representations in Table \ref{irreps} were tried, and the only one leading to a lower residual than using $\Gamma_1$ alone was the combination $\Gamma_1 + \Gamma_2$ (magnetic $R$-factor = 25.1\%).  In this helical ordering, the moment traces out a circle on the $bc$-plane with the propagation direction of the helix along $a$.  This transverse helical structure shown in Fig.~\ref{FeTe_SPINS} would also appear like the transverse SDW of $\Gamma_1$ when projected down the $ab$-plane.  However, the helical model is a better fit to the neutron powder data than the SDW ordering.  The moment size of 1.60(2) $\mu_B$/Fe for both the in-plane and interstitial iron sites indicates an approximate 10 \% decrease from the moment size found in the commensurate magnetic phase.  Adding a spin component along the $a$-direction led to a small moment size ($< 0.2 \mu_B$) with an uncertainty larger than the actual parameter.  We therefore rule out the possibility of a spin-component along the $a$-direction.

The polarized neutron measurements on a crystal with composition Fe$_{1.124(5)}$Te revealed two magnetic propagation vectors in the ground state.  The contour maps of the magnetic scattering versus temperature upon warming and cooling are presented in Fig.~\ref{FeTe_1pER_SPINS}a--d.   As can be seen in the contour maps, the ordering with H = 0.3855(2) has intensity in both the NSF and SF channels while the one at  H = 0.4481(4) only shows intensity in the NSF channel.  Cross sections at different temperatures are presented in Fig.~\ref{FeTe_1pER_SPINS}e--g, where the contribution from the NSF and SF for both peaks is clearly shown.  According to our experimental setup (Fig.~\ref{FeTe_SPINS}), this implies that the ordering for H = 0.3855(2) is non-collinear while that for H = 0.4481(4) is collinear.  In this section, we focus on the incommensurate structure with H$ \approx 0.385$ since this corresponds to the helical ordering, and return to the other in the following section.

From the BT-1 powder data,  the moment of the helical ordering was found to be constrained to the $bc$-plane with equal contributions to the $b$ and $c$ directions.   By analyzing the SF/NSF ratio from the polarized single crystal experiments, the contribution to each axis can be calculated.  Using the lattice parameters from NPD data of a similar composition, the cosine angle between $\mathbf Q = $($\delta \, \, 0 \, \, \frac 1 2$) and $\mathbf c^*$ was calculated to be approximately 51.2$^{\circ}$, which leads to the angle between $\mathbf{c^*}$ and $\mathbf{S_{\perp}}$ to be $38.8^{\circ}$.  Therefore, if the spin amplitude in the $b$-direction and the $c$-direction are equal, the SF/NSF ratio should be equal to $\cos$ ($38.8^{\circ}$).  However, the SF/NSF ratio is less than this value ($\approx 0.78$) and remains constant around 0.5 within error as a function of temperature.  The average value of of 0.49(7) for the SF/NSF ratio leads to a spin amplitude maximum in the $c$-direction that is about 63(9)\% the value in the $b$-direction.  Evidently, in this composition of Fe$_{1.125(5)}$Te the helical structure does not trace out a circle in the $bc$-plane as found for Fe$_{1.143(3)}$Te, but instead an ellipse elongated in the $b$-direction.

\subsection{Short-range spin density wave ordering in $x \approx12\%$}

\begin{figure}[!b] \centering
\includegraphics[width=0.9\linewidth,angle=0.0]{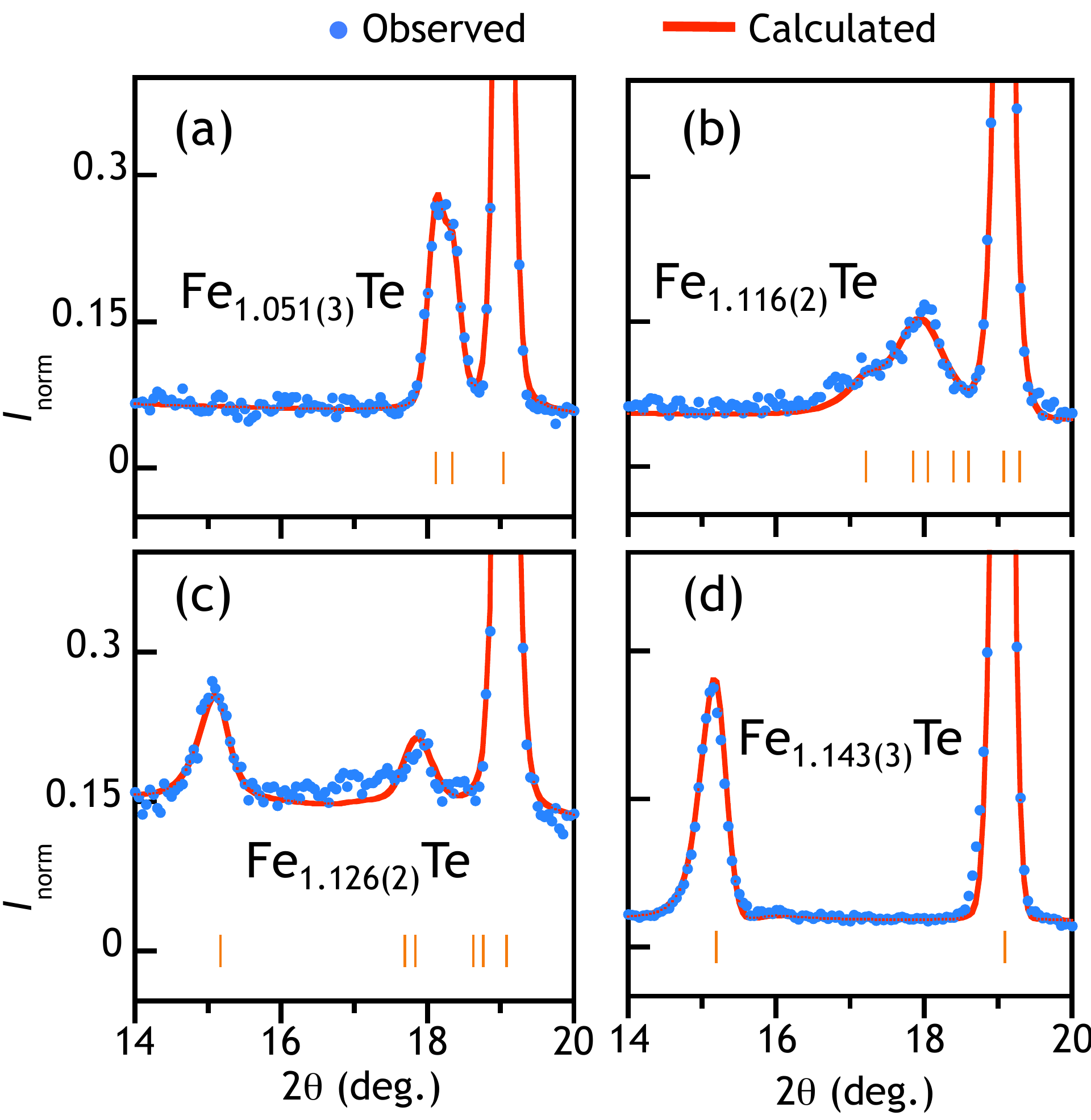}
\caption{[color online]  The evolution of the magnetic propagation vector with interstitial iron as evidenced by the change in the low-angle magnetic scattering from BT-1 neutron data.  All data is normalized to the (001) nuclear Bragg peak at $\approx 19.1$ deg. $2\theta$.  In (a) the monoclinic Fe$_{1.051(3)}$Te has two closely separated magnetic Bragg peaks which correspond to the (+0.5 0 +0.5) and (-0.5 0 +0.5) satellite positions.  In (b) the broad magnetic scattering is fit with two magnetic phases, slightly incommensurate at (0.489(1), 0, 0.5) and  (0.460(1), 0, 0.5).  In (c) the two well separated peaks can be fit with an incommensurate and slightly incommensurate mangetic ordering.  (d) Single incommensurate ordering at (0.380(2), 0, 0.5) for  Fe$_{1.143(3)}$Te.}
\label{FeTe_magnetic}
\end{figure}

From the NPD data, the composition of $x\approx 12\%$ leads to crystallographic phase separation and to short-range magnetic order.   In Fig.~\ref{FeTe_magnetic}a--c  the low-angle magnetic peaks for several compositions shows that as excess iron is increased, the position, intensity, and shape of the magnetic peak changes.   For Fe$_{1.051(3)}$Te the bicollinear commensurate ordering fits the two closely separated magnetic Bragg peaks close to the nuclear (001) peak (Fig.~\ref{FeTe_magnetic}a).  At the other end of the phase diagram, Fe$_{1.143(3)}$Te, the magnetic peak is fit with the incommensurate helical ordering discussed above (Fig.~\ref{FeTe_magnetic}d) .  For compositions near $x \approx 12 \%$, however, the magnetic scattering does not correspond to either the commensurate bicollinear structure nor the incommensurate helical structure.  Furthermore, the peaks in this composition are significantly broadened with respect to the nuclear peaks, indicating these are not long-range ordered magnetic structures.  For Fe$_{1.119(1)}$Te, the peaks move to an intermediate scattering angle and appear to consist of two magnetic phases (Fig.~\ref{FeTe_magnetic}b), which is consistent with the observation of two crystallographic phases at base temperature.  The trend continues for Fe$_{1.126(2)}$Te, where the magnetic phases are separate enough to be distinguishable, and the structure also consists of two crystallographic phases.

While we can fit the magnetic structures of Fe$_{1.051(3)}$Te and Fe$_{1.143(3)}$Te satisfactorily with the BT-1 powder data, the fits to the intermediate structures of  Fe$_{1.119(1)}$Te and Fe$_{1.126(2)}$Te are complicated by their multi-phase compositions.   Therefore, we fit a combination of incommensurate structures (representation $\Gamma_1$) only to obtain the propagation vector of these intermediate phases (Fig.~\ref{FeTe_magnetic}b,c).  The propagation vectors \propinc were found to have $\delta = 0.490$(1) and $\delta = 0.461$(1)for Fe$_{1.119(1)}$Te and $\delta = 0.483$(1) and $\delta = 0.379$(1) for Fe$_{1.126(2)}$Te.  The limited information from powder diffraction data, however, make it impossible to distinguish whether this short-range order magnetic phase corresponds to an SDW or a helical ordering.   The polarized neutron diffraction studies on a single crystal sample clarifies this ambiguity.

For the crystal with composition Fe$_{1.124(5)}$Te the contour maps of the magnetic scattering from the SPINS data versus temperature are presented in Fig.~\ref{FeTe_1pER_SPINS}a--d.   The presence of two propagation vectors is consistent with the powder studies showing that for  $x \approx 12\%$, a short-range magnetic ordering appears along with crystallographic phase separation.  The coexistence of two incommensurate structures in a single sample allows us to compare the difference between the incommensurate helical ordering with H = 0.3855(2) and the intermediate one with H = 0.4481(4) simultaneously.   As can be seen in the contour maps, the feature at H = 0.4481(4) only has intensity in the NSF channel, and its peak width is considerably broadened compared to that at H = 0.3855(2), which is resolution limited (Fig.~\ref{FeTe_1pER_SPINS}e--g).  Therefore, we can conclude that the broad, slightly incommensurate magnetic structure observed in both powder and single crystal samples corresponds to a short-range ordered SDW.

\begin{figure}[!b] \centering
\includegraphics[width=0.9\linewidth,angle=0.0]{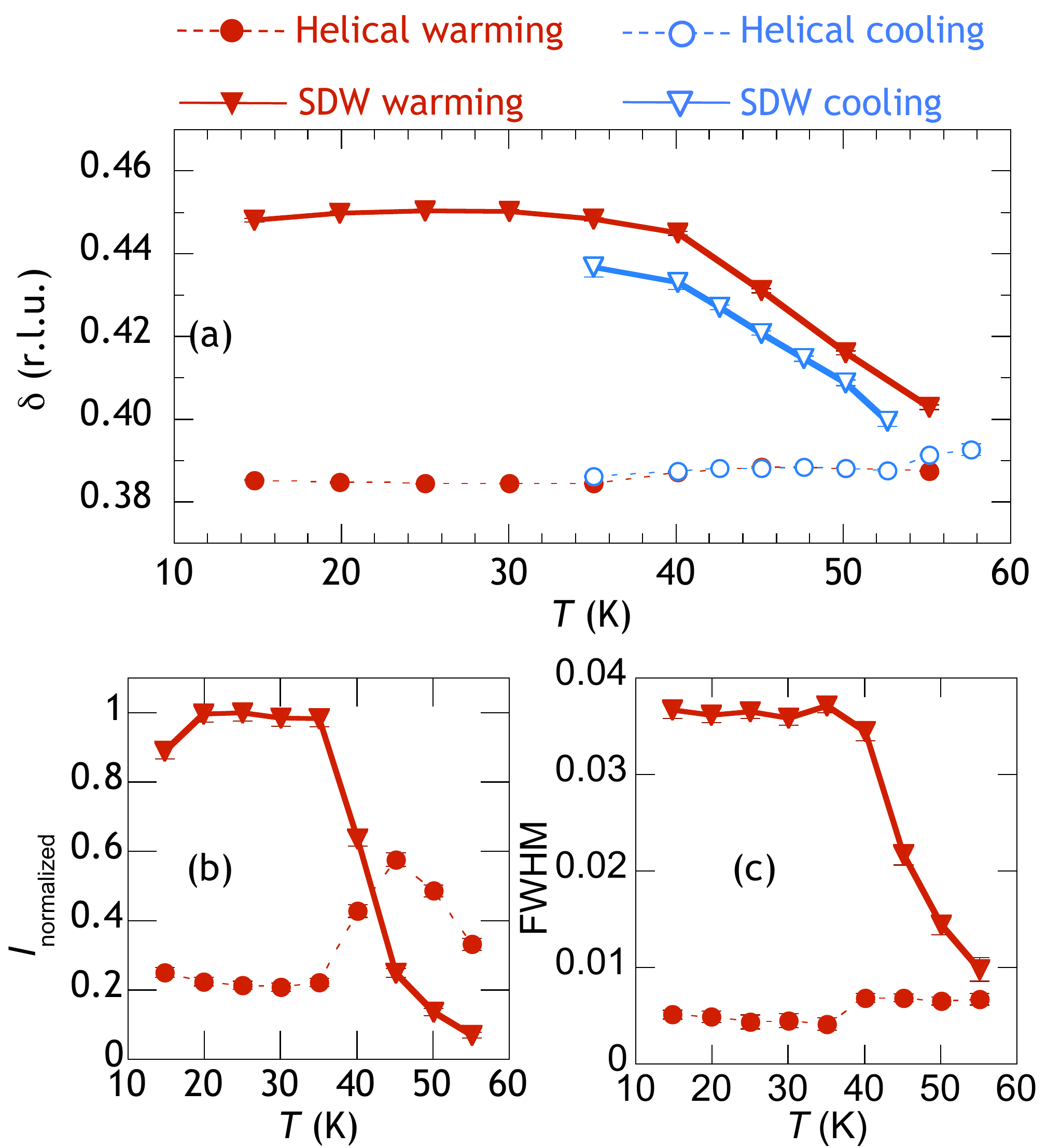}
\caption{[color online]  Peak centers and integrated intensities from fits to the magnetic Bragg peaks of a single crystal Fe$_{1.124(5)}$Te measured in the spin polarized experiments on the SPINS spectrometer.  In (a) the peak centers and therefore the $\delta$ from the propagation vector \propinc is shown versus temperature upon warming and cooling.  In (b) the integrated intensities for the Fe$_{1.124(5)}$Te crystal upon warming. In (c) the full width at half maximum of the two competing magnetic peaks is shown upon warming.}
\label{FeTe_SPINS_analysis2}
\end{figure}

Although the two magnetic phases could be due to a heterogenous distribution of interstitial iron, the temperature behavior of the two magnetic peaks suggest otherwise as the two phases interact with each other.   As the N\'eel point is approached upon warming, the propagation vector of the SDW moves towards that of the helical structure found at H = 0.4481(4) (Fig.~\ref{FeTe_SPINS_analysis2})a.  Furthermore, the spectral weight of the short-range SDW shifts to the helical structure upon warming as shown in Fig.~\ref{FeTe_SPINS_analysis2}b, where the integrated intensities of the peaks are plotted versus temperature.  At base temperature the short-range SDW comprises most of the spectral weight; above 35 K, the intensity diminishes linearly for the SDW while that of the helical structure increases.  Since the total magnetic intensity eventually declines, this produces a maximum in the intensity of the helical magnetic ordering around 45 K (Fig.~\ref{FeTe_SPINS_analysis2}b).

At base temperature the peak of the incommensurate SDW in Fe$_{1.124(5)}$Te is much broader than the incommensurate helical structure, as was observed in the powder diffraction patterns (Fig.~\ref{FeTe_magnetic}b--d).  Upon warming, the full-width at half maximum (FWHM) decreases linearly above 35 K (Fig.~\ref{FeTe_SPINS_analysis2}c), which corresponds to the same temperature at which the peak starts to lose intensity.  Eventually, the FWHM of the SDW becomes nearly equivalent to that of the helical structure.  Apparently, the SDW is becoming long-range ordered with increasing temperature as its propagation vector moves from $\delta \approx 0.45$ to $\delta \approx 0.40$.  Nevertheless, it remains an SDW with its moments along $b$-direction as shown by comparing the NSF and SF scattering (Fig.~\ref{FeTe_1pER_SPINS}e--g).

\begin{figure}[!t] \centering
\includegraphics[width=0.9\linewidth,angle=0.0]{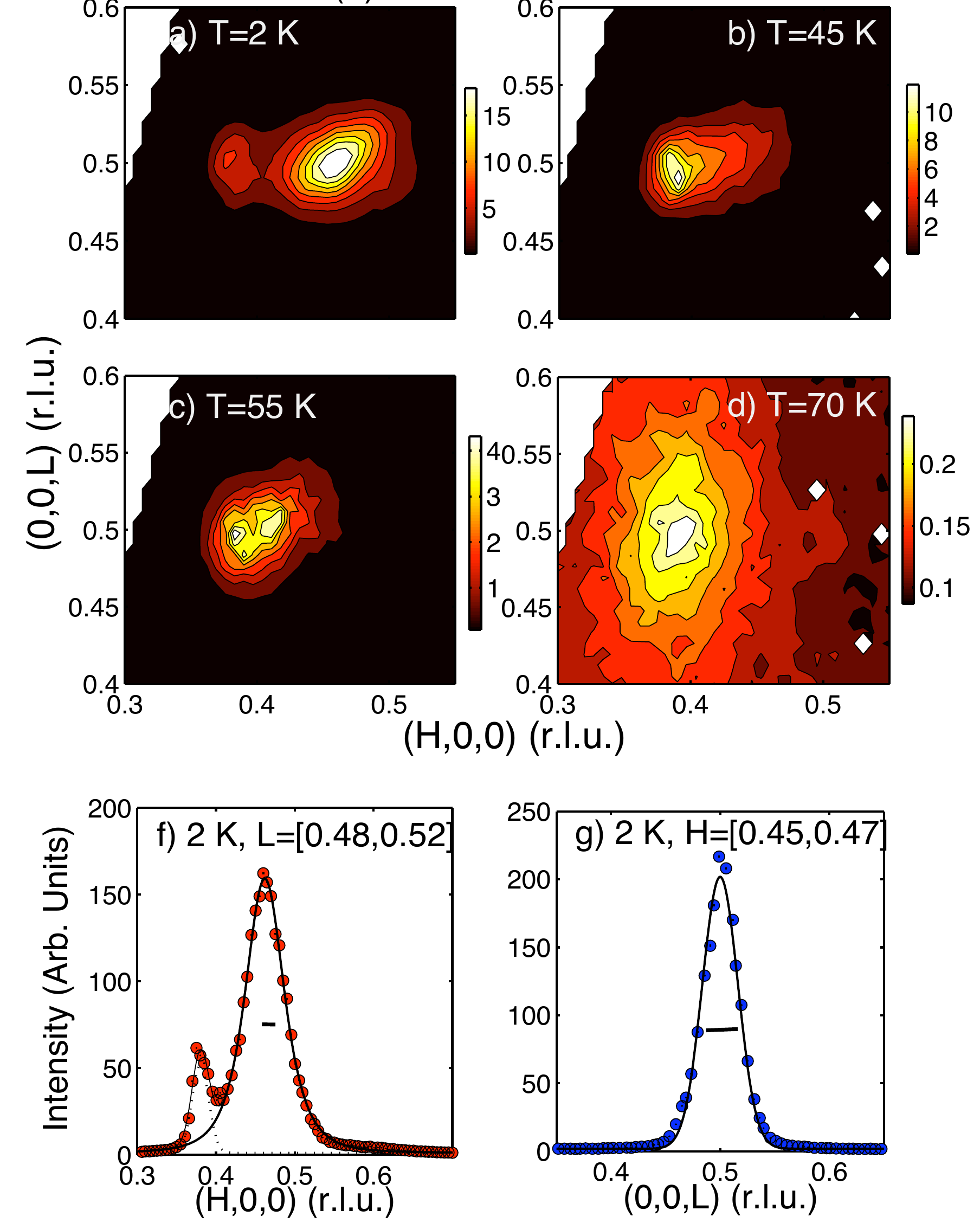}
\caption{[color online]  Contour maps of the (H 0 L) plane of Fe$_{1.124(5)}$Te taken on the MACS spectrometer.  The final and incident energies were set to 3.6 meV, to capture only the elastic scattering of the two magnetic peaks.  The change in peak position and intensity as a function of temperature is shown in panels (a)--(d).  In (f) the peaks are integrated in L over a range of 0.48 to 0.52, which shows that the broad short-range magnetic scattering centered at H = 0.45 dominates most of the scattering at base temperature.  In (g) the width of the peak, after integrating in H over 0.45 to 0.47, shows that along the L-direction, the magnetic ordering is long-range.  The bars inside the peaks represent the instrument resolution.}
\label{FeTe_MACS}
\end{figure}

In the SPINS experiment, the peaks were measured only along (H 0 $\frac 1 2$), so that information on their profile in the L-direction is unknown.  This loss of information means that the linear change in the intensity and width with temperature could be due to shifts of the peak in the L-direction and not to an actual change in the peak shape.  Therefore, we performed two-dimensional scans of the magnetic scattering to ensure that the SDW is shifting spectral weight to the helical ordering while becoming long-range ordered.

The MACS spectrometer is optimized for constructing two-dimensional maps in reciprocal space at different energy transfers.  We performed measurements of the (H 0 L) plane on the same crystal from the SPINS experiment at zero energy transfer to obtain the temperature behavior of the purely static magnetic ordering.  As shown in Fig.~\ref{FeTe_MACS}a, the (H 0 L) map reveals that the incommensurate SDW is quite broad in H.  At base temperature, integrating over a range of L = 0.48 to 0.52, leads to the intensities of the mangetic peaks shown in Fig.~\ref{FeTe_MACS}f, which differ from those of the SPINS experiment (Fig.~\ref{FeTe_1pER_SPINS}e).  This is due to the fact that in SPINS, the scan is only along (H$ \,\,0\,\, \frac 1 2$), whereas in MACS one can integrate the intensities of the magnetic peaks over a wide L-range.  Clearly the short-range SDW magnetic structure dominates the scattering at base temperature in this composition.

As the temperature is increased up to the N\'eel point, the intensity shifts dramatically towards the H$ = 0.385$ peak (Fig.~\ref{FeTe_MACS}a--c).  Above the ordering temperature, the scattering has become quite diffuse with significant broadening in L (Fig.~\ref{FeTe_MACS}d).  The correlation length along H of the short-range SDW can be found by fitting a Lorentzian squared term to the peak shown in Fig.~\ref{FeTe_MACS}f.\cite{Birgeneau_1983}  The correlation length along H was found to be $22(3)$ \AA~for the SDW in Fe$_{1.124(5)}$Te at base temperature.  The integrated intensity of this peak along (0 0 L) shows that it is actually long-range ordered in this direction as the peak width was nearly equal to the instrument resolution (Fig.~\ref{FeTe_MACS}g).

Overall, the MACS data corroborates the SPINS data, and both suggest that the two magnetic ordering vectors within the same crystal ($x \approx 12\%$) occur due to some electronic phase separation and not microscopic chemical phase separation.   Furthermore, this crystal was shown by single crystal XRD to be a single phase at 250 K.  Likewise for the NPD patterns of Fe$_{1.126(2)}$Te above the structural transition, only a tetragonal phase is sufficient to describe the structure.  For the commensurate phase Fe$_{1.09(1)}$Te, the SDW makes a brief appearance only close to the N\'eel point and is long-range.   This suggests that in order to observe the short-range SDW, electronic phase separation has to occur, which would also lead to the observed structural phase separation at base temperature.  This structural frustration leads to a freezing in of the short-range SDW.

\section{Discussion and Conclusions}

To summarize some of the key results presented above, we have constructed a magnetic-crystallographic phase diagram of our 13 powder and 6 single crystal samples.   The $\delta$ of the propagation vector \propinc is plotted against the interstitial iron concentration; various significant information regarding the magnetic and crystallographic structure is also included (Fig.~\ref{FeTe_phases}).  What is apparent from this diagram is that the tuning of the propagation vector according to interstitial iron does not vary linearly as proposed in an earlier study.\cite{Bao_2009}   Instead, there is a threshold of interstitial iron required to change the propagation vector from \propcom to \propinc.  At this critical concentration of $\approx$ 12\%, crystallographic phase separation including orthorhombic and monoclinic symmetries occurs down to base temperature.  Within this phase an incommensurate and short-range ordered spin-density wave (SDW) freezes in.  Previously unknown for Fe$_{1+x}$Te, this SDW order appears at $\mathbf k =$ ($\approx 0.45$ 0  $\frac 1 2$) at base temperature and upon warming, changes its position in the H-direction, becomes long-range ordered, and shifts its spectral weight to the other incommensurate ordering.  It is also important to note that the bicollinear commensurate phase also has an SDW that appears at higher temperatures close to the N\'eel point and competes with the commensurate phase.  

Constant energy scans of the (H 0 L) plane of a single crystal reveal that this SDW has a correlation length of 22(3) \AA~in the H-direction, but is long-range ordered in the L-direction.  As more interstitial iron is added, a spin component in the $c$-direction develops so that the SDW structure gives way to a helical structure elongated in the $b$-direction (an elliptical helix).   Once at $x = 14.3(3)\%$, the spin-components in the $b$ and $c$-directions are equal, so that the structure can be described as a circular helix with a turn angle of $\pi \delta$ ($\approx 69.3 ^{\circ}$).  This study shows that the magnetic phase diagram of Fe$_{1+x}$Te is much richer than initially suspected, and all of these structures are illustrated in Fig.~\ref{FeTe_phases}.

\begin{figure}[!b]\centering
\includegraphics[width=0.9\linewidth,angle=0.0]{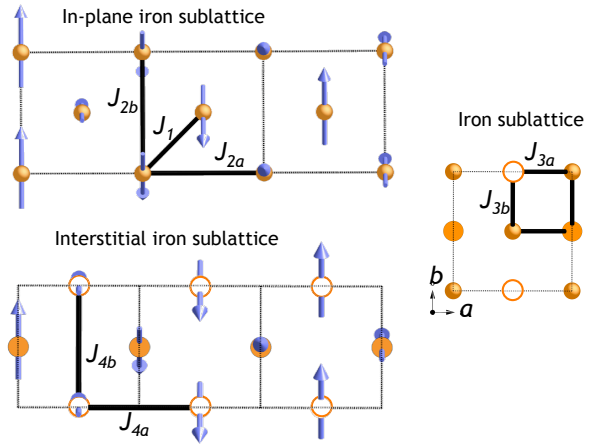}
\caption{[color online]  Labeling of the exchange parameters for the helical ordering present in Fe$_{1.143(5)}$Te.  On top, the three exchange parameters used for the in-plane iron atoms (rendered) .   Similar exchange parameters are present for the interstitial sites, which are shown below with filled circles representing the site at $z \approx 0.7 $ and the empty circles the site at $z \approx 0.3 $.  The exchange parameters linking the two iron sublattices are shown to the right.}
\label{FeTe_exchange}
\end{figure}

We can attempt to explain the variety of observed magnetic structures to a first approximation within a local moment picture.   In the helical ordered state, we label the exchange parameters between the nearest neighbor iron cations as $J_1$ and second-nearest neighbor cations as $J_2$, which is split into $J_{2a}$ and $J_{2b}$ due to the orthorhombic distortion.  For the interstitial iron sites, the same interactions are relevant and are labeled $J_3$ for nearest neighbor interactions with the in-plane iron sublattice, and $J_4$ for the nearest neighbor interaction among the interstitial sites.  Again, the orthorhombic distortion causes $J_3$ to split into $J_{3a}$ and $J_{3b}$. The labeling of the exchange parameters in shown in Fig.~\ref{FeTe_exchange}.  Based on the known helical structure we found for Fe$_{1.146(3)}$, we can write an expression relating the exchange parameters using the classical Heisenberg formulation.

\begin{eqnarray}
H  & =  & \sum_{<i,j>} J_{i,j} \mathbf S_i \cdot \mathbf S_j \\
E  & =  & N_1S^2[J_1 \cos(\alpha) + J_{2a} \cos (2\alpha) + J_{2b}] \nonumber \\
{}  & {}  & + N_2S^2[J_{3a} \cos (\alpha) +  J_{4} \cos (2\alpha) + J_{3b}]
\end{eqnarray}

where $\alpha$ is the turn angle along the $a$-direction ($= \pi \delta  \approx 69.3^{\circ}$), $N_1$ is the number of nearest neighbor and second-nearest neighbors within the in-plane iron square lattice and $N_2$ is the number of nearest and second nearest neighbors between the interstitial iron and other sites.   We can then find the relation between the exchange parameters that would lead to the lowest energy with respect to the turn angle.

\begin{eqnarray}
\frac {dE}{d\alpha} & = & -N_1S^2[J_1\sin(\alpha) - 2J_{2a}\sin(2\alpha)] \nonumber\\
{}  &  {}  & -N_2S^2[J_{3a}\sin(\alpha) - 2J_4\sin(2\alpha)] \\
0 & = &  -(N_1J_1 + N_2J_{3a})\sin(\alpha) \nonumber\\
{} & {} &  - 4(N_1J_{2a}+N_2J_4)\sin(\alpha)\cos(\alpha)
\end{eqnarray}

\begin{equation}
\frac {(N_1J_1  +  N_2J_{3a})} {(N_1J_{2a} + N_2J_4)}   =  -4\cos{\alpha} \\
\end{equation}

This localized model shows that for the observed propagation vector ($4 \cos \alpha > 1$), the nearest neighbor interactions become greater than the second-nearest interactions.  Therefore, the helical structure is a result of frustration in $J_1$ and $J_{3a}$, which becomes greater as the number of interstitial iron sites $N_2$ become populated.

For the other extreme of the phase diagram, the commensurate bicollinear phase, the monoclinic setting splits $J_1$ into  $J_{1a}$ and $J_{1b}$, relieving the frustration present in the orthorhombic phase.  Inelastic neutron scattering measurements of the spin waves in a crystal of low interstitial iron, found that  $J_{1a}$ and $J_{1b}$ are indeed highly anisotropic and ferromagnetic whereas $J_2$ was found to be smaller, antiferromagnetic, and isotropic.\cite{Lipscombe_2011}  This implies that the denominator of Eqn. (5) plays less of a role determining the observed changes in the magnetic ordering in Fe$_{1+x}$Te.  

From Eqn. (5), one can calculate what the angle and therefore propagation vector should be for the case of the ratio of the exchange parameters becoming equal.  This propagation vector corresponds to $\delta = 0.42$, which is incidentally equal to the propagation vector of the SDW competing with the commensurate phase in Fe$_{1.09(1)}$Te (Fig.~\ref{FeTe_comm_SPINS}) as the N\'eel point is approached.  This suggests that as the structure is getting closer to the tetragonal setting, the nearest and next nearest neighbor interactions become approximately equal in magnitude.

From our NPD results, there are no obvious changes in the $ab$ plane as interstitial iron is increased.  The largest changes occur in the $c$ parameter, which decreases as $x$ increases and the monoclinic angle $\beta$, which widens as $x$ increases.  Instead of the entire crystal changing to the orthorhombic phase, however, phase separation occurs.  This phase separation seems to be necessary to observe SDW ordering with a short-range correlation length.  We speculate that an interplay between $J_{1}$ and $J_{3a}$ and a strong anisotropy in the $b$-direction leads to a collinear SDW and then an elliptical helix elongated along the $b$-direction, rather than a circular helical ordering.

More recent inelastic neutron scattering measurements that probe low-energy excitations of two crystals representing opposite ends of the phase diagram shed light on the role of interstitial iron on the magnetic fluctuations in Fe$_{1+x}$Te.\cite{Stock_2011}   The dispersion curves of the magnetic excitations have revealed that the commensurate phase with low interstitial iron Fe$_{1.057(7)}$Te has an energy gap of 7 meV at base temperature.  The other extreme of the phase has no such energy gap, but is peaked at an energy of around 4 meV.  This suggests that in the commensurate phase there exists an anisotropy gap favoring moments aligned along $b$.  This energy gap is incidently of the same magnitude as the spin resonance observed in the superconducting phases.  The fact that the crystal with stoichiometry Fe$_{1.141(5)}$Te has no gap is perfectly consistent with our diffraction studies showing a helical ordering with moments directed towards the $b$- and $c$-direction, \textit{i.e.} the anisotropy gap is closed.

Several theoretical studies have tried to answer the question of what microscopic mechanisms are responsible for the exchange interactions driving the observed ordering of Fe$_{1+x}$Te to be different from the rest of the parent phases of iron-based superconductors.  The different approaches can be broadly summarized into three different models: the itinerant picture predicting the observed ordering based of Fermi surface nesting, the localized moment model, and finally the orbital-ordered and double exchange model.

The localized model nicely captures a lot of the features of the commensurate and helical phase as explained above.  Indeed, the first-principles electronic structure calculations by Ma \textit{et al.} reproduces the bicollinear ordering by including nearest-, second nearest-, and third nearest-neighbor interactions within the Heisenberg model.\cite{Ma_2009}  This $J_1$-$J_2$-$J_3$ model concludes that the dominating exchange parameter is $J_2$ (next nearest) and that these interactions arise primarily from superexchange with the Te $5p$ orbitals acting as the mediating states.   These results are, however, inconsistent with the neutron scattering work of Lipscombe \textit{et al.}, which show that $J_1$ is not antiferromagnetic and that $J_2$ is not greater than $J_1$.\cite{Lipscombe_2011}  The study by Fang \textit{et al.} also address the magnetism within a localized model and find that a critical amount of interstitial iron induces incommensurate ordering mostly by affecting the strong coupling of lattice and magnetic degrees of freedom in Fe$_{1+x}$Te.\cite{Fang_2009}

In the itinerant model, Zhang \textit{et al.} have performed density functional studies to show that the interstitial iron in Fe$_{1+x}$Te acts as a strong local moment that interacts with the itinerant magnetism of the in-plane iron.\cite{Zhang_2009}   They model the interstitial iron as a Fe$^+$ site within a supercell, so that they study an interstitial iron concentration of 12.5\%, and this moment is enough to drive Fe$_{1+x}$Te to have the observed bicollinear ordering as opposed to a ferromagnetic, checkerboard antiferromagnetic, or nonmangetic arrangement.  

A later electronic structure study by Han and Savrasov calculates  the same Fermi surface nesting vector ($\pi$, $\pi$) in Fe$_{1+x}$Te as in the iron arsenides.  This difference between their calculated structure and that observed is explained as arising from the fact that Fe$_{1+x}$Te is self-doped by the electrons of the interstitial iron site.\cite{Han_2009}  In their calculations, this is enough to reshape the Fermi surface and lead to the observed ($\pi$,0) nesting, which corresponds to the bicollinear magnetic structure.  Han and Savrasov have to make some unphysical assumptions, however, such as the interstitial iron donating all of its valence electrons and therefore having an $8^+$ oxidation state.  We know from the diffraction data presented in this paper that the interstitial iron has a moment equal to that of the in-plane iron, which implies that it also has an oxidation state of $2^+$.

Finally, the orbital ordering picture by Turner \textit{et al.} offers an interesting model to explain the observed magnetic ordering.\cite{Turner_2009}  In this model, correlated local moments and orbital degeneracy lead to a strong anisotropy towards the $b$-direction.  Furthermore, the structural distortion leading to the orthorhombic cell is proposed to arise from orbital ordering rather than magnetic ordering.  Double exchange leads to the ferromagnetic aligning in the $b$-direction, and a kinetic energy term leads to antiferromagnetic coupling in the $a$-direction.  Electron doping from the interstitial iron leads to further occupation of the orbital responsible for ferromagnetic coupling, and the incommensurate spiral structure arises from the system trying to lower the energy from the nearest neighbor interaction (J$_1$ in our model).  Interestingly, this study also predicts phase separation in the incommensurate ordering with the doped electrons separating into high- and low-density regions.   This is an appealing model to help explain the phase separation observed in our samples for $x \approx 12\%$ since the temperature behavior of the magnetic scattering implies that the phase separation is electronic in nature.

The SDW observed in the phases Fe$_{1+x}$Te for $x \approx 12\%$ could help explain some of the features observed in the nonsuperconducting Se-doped phases.   In several studies undertaken to understand why the ($\pi$, 0) ordering gives way to the spin resonance with ($\pi$, $\pi$) symmetry, several studies have revealed a slightly incommensurate ordering that it static and short-range ordered.\cite{Xu_2010}  In some samples, this short-range ordering with a weak moment for iron ($\approx 0.1 \mu_B$) even seems to coexist with superconductivity.  Interestingly the propagation vector of $\mathbf k \approx $ (0.45\, 0\, $\frac 1 2$) found in some samples,\cite{Wen_2009} is close to that found in the SDW presented in this paper.  

Similar studies on lightly doped samples of Fe$_{1+x}$Te$_{1-y}$Se$_y$ show that the incommensurate short-range magnetic ordering crosses over from static to purely dynamic as a function of Se-doping.  Through a combination of inelastic neutron scattering measurements and magnetization measurements, Katayama \textit{et al.} discovered a spin-glass transition that seems to compete with long-range ordering.\cite{Katayama_2010}  Interestingly, the SDW is also centered at a propagation vector of ($0.46$, 0, $\frac 1 2$), very close to that of the SDW found for Fe$_{1+x}$Te in this paper.  Katayama \textit{et al.} also find evidence for crystallographic distortion, which would be consistent with our observation that the short-range SDW was only found in samples of with crystallographic phase separation.

The similarities between our findings and those for Se-doped samples implies that the critical amount of interstitial iron plays a similar role to that of Se-doping in modulating the bicollinear antiferromagnetic structure of Fe$_{1+x}$Te.  The interstitial iron does this by increasing the value of the nearest neighbor exchange interactions.  Conversely, the Se substitution affects the nearest neighbor interaction but seems to suppress rather than increase this exchange interaction.  Thus, in the superconducting phases, the Se-doping destroys long-range magnetic ordering altogether while in the case of Fe$_{1+x}$Te, interstitial iron doping causes the long-range bicollinear ordering to be replaced by other types of lower energy including the incommensurate short-range SDW and the incommensurate long-range helical ordering.   

A significant difference between the two types of doping (interstitial iron vs. anion substitution) is that the excess iron causes only short-range correlations along H, while Se-doping leads to short-range correlations along L.  Thus, Fe$_{1+x}$Te never becomes a two-dimensional magnet within the $ab$-plane.  Indeed, this is the opposite of what what has been observed in some cuprate phases such as YBa$_2$Cu$_3$O$_{6.353}$ and La$_{2-x}$Sr$_x$CuO$_4$, which have short-range correlations along the interplanar spacing and long-range within the planes.\cite{Stock_2008, Stock_2006, Fujita_2002}  Likewise, when the superconducting state is reached in Fe$_{1+x}$Te$_{1-y}$Se$_y$ the position of the spin resonance in reciprocal space from inelastic neutron studies implies that it becomes strongly two-dimensional in the $ab$-plane.  Since the interstitial iron chemically connects the layers, it could hinder superconductivity by maintaing long-range magnetic ordering in the L-direction.  Indeed, this deleterious effect of interstitial iron on superconductivity has been observed in several studies.\cite{Rodriguez_2010b, Liu_2009, Viennois_2010}

\section{Acknowledgements}
This work has benefited from the use of the NPDF beamline at the Lujan Center at Los Alamos Neutron Science Center, funded by the US DOE Office of Basic Energy Sciences.

%--------------------------------------------------------------------------------------------------------------------------------------------------------
%  Bibliography
%
%\bibliographystyle{apsrev}
%\bibliography{FeTe_1pER}

\end{document}